\documentclass[preprint]{aastex}

\slugcomment{Submitted to: The Astrophysical Journal}

\shorttitle{Lensing of Slow-Moving Particles}
\shortauthors{Patla et al.}

\begin{document}


\title{Flux Enhancement of Slow-moving Particles by Sun or Jupiter: Can they be Detected on Earth?}


\author{Bijunath R. Patla}
\email{bpatla@cfa.harvard.edu}
 \affil{Smithsonian Astrophysical Observatory,
 Harvard-Smithsonian Center for Astrophysics,\\
60 Garden St, Cambridge, MA 02138, USA}

 \author{Robert J. Nemiroff}
 \affil{Department of Physics, Michigan Technological University,\\
 1400 Townsend Dr.,
Houghton, MI, 49931,USA}

\author{Dieter H. H. Hoffmann}
 \affil{Technische Universit\"{a}t Darmstadt, Institut f\"{u}r Kernphysik,\\
  Schlossgartenstr. 9, 64289 Darmstadt, Germany}

\and

\author{Konstantin Zioutas}
\affil{University of Patras, GR-26500 Patras, Greece}

\date{\centering{\today}}



\begin{abstract}
Slow-moving particles capable of interacting solely with gravity
might be detected on Earth as a result of the gravitational lensing induced focusing action of the Sun. The deflection experienced by these particles are inversely proportional to the square of their velocities and as a result their focal lengths will be shorter. 
We investigate the velocity dispersion of these slow-moving particles, originating from distant point-like sources, for imposing upper and lower bounds on the velocities of such particles in order for them to be focused onto Earth. 
We find that fluxes of such slow-moving and non-interacting  particles must have speeds between $\sim 0.01 $ and $.14 $ times the speed of light, $c$. Particles with speeds less than $\sim 0.01 c$ will undergo way too much deflection to be focused, although such individual particles could be detected. At the caustics, the  magnification  factor could be as high as $\sim 10^{6}$. 

We impose lensing constraints on the mass of these particles in order for them to be detected with large flux enhancements to be greater than $10^{-9}$~eV. An approximate mass density profile for Jupiter is used to constrain particle velocities for lensing by Jupiter. We show that Jupiter could potentially focus particles with speeds as low as $\sim0.001c$, which the Sun cannot.  
As a special case, the perfect alignment of the planet Jupiter with the Sun is also considered.
\end{abstract}






\section{Introduction}
According to the estimates using latest data, $80\%$  of the matter content of the universe, most likely comprises non-baryonic dark matter particles that are  yet to be discovered~\citep{nasa,planck}. Some or most of these particles could be heavy and hence slow-moving. Furthermore, if they do not interact with any of the known force fields but  gravity, they are capable of 
passing through (transparent to) the Sun's interior~\citep{lawrence,clark,ohanian,bontz, burke,gerver,demkov,dhh,patla08}.

The velocities of massive particles, in order to be consistent with the theory of special relativity, must be smaller than the speed of light in vacuum,$c$, and therefore, will undergo considerably larger deflection for the same impact radius compared to photons traveling in a gravitational field or even neutrinos that are often considered to 
be fast moving ($v \approx c$).

For photons and neutrinos, the minimum focal 
length of the Sun is about 550~AU and 23.5~AU respectively.
The magnification of the particle flux from point-like distant sources can be detected only by observers
 stationed 
beyond the minimum focus. Hereafter in this paper, when we make references to slow-moving 
particles, it is to be assumed that they interact only with gravity.

For slow-moving particles the minimum focal length is shorter because, from elementary projectile motion, the deflection angle, $\delta\propto v^{-2} b^{-1}$, where $v$ is the speed of the particle impacting the lens at a distance $b$ from the center. In this paper we seek to find answers for the following questions: 1) What is the velocity range for particles in order for them to be focused at 1~AU ? 2) What is  the requirement on the
minimum mass of particles for it to be focused at 1~AU  and also record an appreciable flux (magnification) without getting scattered due to diffraction effects? 3) What is the magnitude of the maximum amplification? 4) What is the field of view within which a source has to be for the flux of particles emitted by it to be amplified by large factors? 5) Can the flux density and velocity distributions of these particles be used to further our understanding of fundamental physics, say, for example, testing the Equivalence Principle (or its violation)? 

A detection will be a direct confirmation of the existence of non ordinary matter.
We use the appropriate deflection formula in the weak field limit for slow-moving particles to put limits on their velocities in order for them to be detected at 1~AU. We are building on the previous work done by~\citet{dhh} using the formalism developed in~\citet{patla08}.

The plan of the paper is as follows: \S 2 summarizes particles that are considered to be slow-moving. Section 3 discusses two formulas for deflection angles that are used as inputs for solving the lens equation. Section 4 addresses the relationship between focal lengths and impacting velocities, by referring to the previous work of ~\cite{patla08}.  Magnifications and flux density are discussed in \S5. Diffraction limits and mass constraints on the slow-moving particles are detailed in \S6. Lensing by Jupiter and the effects of transiting planets is addressed in \S7, followed by conclusions in \S8.

\section{\label{sec0} Slow-moving particles: WIMPS, Axions}

Plausible candidates for slow-moving particles are classes of Weakly Interacting Massive Particles~(WIMPS) and axions---either or both of which may or may not eventually be confirmed experimentally to match with the observed dark matter density in the universe. 
For excellent reviews on the topic of dark matter, we refer the reader to~\citet{trimble,spergel,drees}. For axions in particular, see~\citet{turner}, \citet{peccei06},~\citet{ringwald}. Surveys looking for clumped baryonic dark matter candidates have, thus far, not been able to confirm their existence~\citep{alcock}.

WIMPs were motivated by the physics of the early, and therefore hot, universe when the conditions  favored reversible processes such as the annihilation of electron-positron pairs.
WIMPs would interact through the exchange of heavy intermediate bosons: $W$, $Z$, higgs.
The extension of the standard model by using the symmetries of the Poincare group, now dubbed supersymmetry~(SUSY), also points to the existence of WIMPs~\citep{diehl,jungman}.
Although recent results from the latest LHC runs are not very encouraging for SUSY~\citep{lhc}. The predicted energies of these WIMPs range from a few GeVs to hundreds of GeVs~\citep{ackermann,picozza,aguilar}. 

During the early 1970's, in order to explain the apparent lack of $U(1)$ symmetry in QCD,  a phase parameter was introduced. This parameter added an extra term to the QCD Lagrangian that violated Parity and Time~(PT) reversal invariance while conserving Charge conjugation(C) invariance, hence violating CP. \citet{peccei77} proposed a solution by introducing a CP-conserving dynamical pseudoscalar field with an effective potential, which when expanded at the minimum yields a massive particle called axion~\citep{weinberg, wilczek}. More recently, axion-like Kaluza-Klein excitations, in the context of higher-dimensional theories of gravity, emergent from stellar interiors have also been a topic of some interest~\citep{dienes, lella}.

Axions that were initially proposed were strongly interacting and massive, but ruled out by experiments later. Flavors of axions that have survived (without being ruled out by experiments) thus far are the so-called invisible axions~\citep{kim,shifman,dine}. These are hypothesized to have formed in the early universe due to spontaneous decay of cosmic strings or by the relaxation of string-domain wall boundaries. In the present epoch axions could seek shelter within the interiors of massive stars or galaxy centers. The  rest masses of these axions fall in the range of $10^{-6}-0.1$~eV, with large error bars on both ends~\citep{raffelt12, hewett}. These bounds are obtained by imposing constraints of energy loss in stars without  affecting stellar evolution or structure formation in the Universe. In particular, for the Peccei-Quinn scale, the WMAP data sets an upper bound of 1~eV for axion rest mass, while a lower bound of $\sim 10^{-6}$~eV follows arguments of overclosure for the Universe.

Models of dark matter flows~(wakes) in the vicinity of gravitating objects have been discussed previously: solar wakes~\citep{solarwakes}, cosmic strings wakes~\citep{stebbins}. Lensing due to dark matter caustics is addressed in~\citep{charmousis}.  In this paper we do not consider these effects. We only consider gravitational lensing of cosmological (located far away from Earth's orbit) point-like sources, that emit slow-moving particles, by the Sun. Isotropic backgrounds would not produce any discernible amplifications.

\section{\label{sec1} Deflection angles for slow-moving particles}
The formula for the deflection angle for a ray of light in the weak field limit is~\citep{einstein36}
\begin{equation}
\delta=\frac{4 G M}{b c^2},
\label{def}
\end{equation}
where $G$ is the gravitational constant, $M$ is the mass enclosed within the impact parameter (radius of cylinder) $b$ and $c$ is the speed of light in vacuum. The value for the deflection at the Solar limb were experimentally confirmed for the first time for visible light and cosmic microwave background radiation by \citet{dyson} and \citet{fomalont} respectively.

\begin{figure}[htb]
\begin{center}
\includegraphics[width=\textwidth]{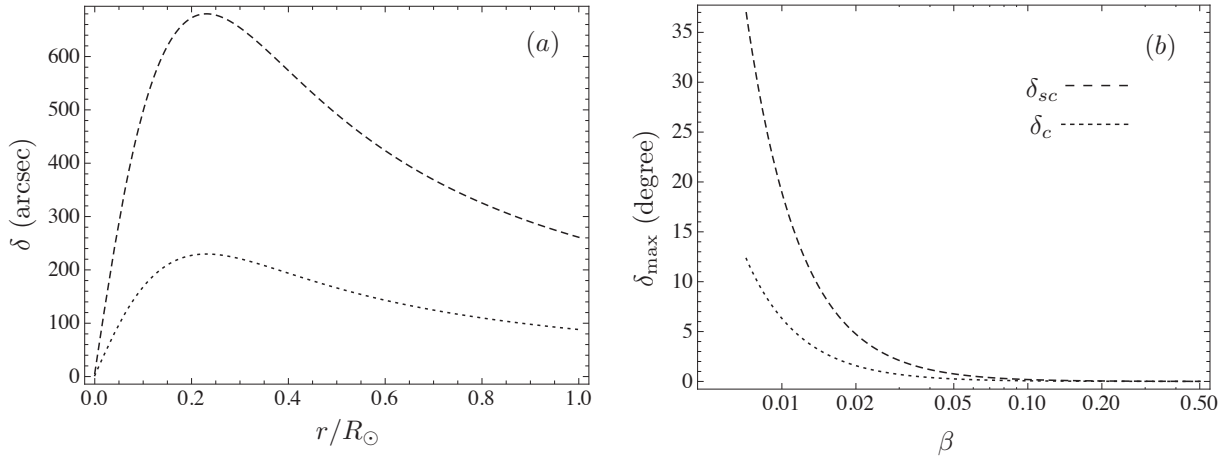}
\caption{Deflection angles for the Sun from SSM data using equations~(\ref{class-def}) and (\ref{sc-def}): (a) as a function of normalized radius of the Sun,(b) maximum deflection angles as a function of particle speed $\beta$. As the particle speed approaches $\sim 0.01 c$ the regnant deflection formulas for both semi-classical and classical formalism yield values that are exceedingly large in order to be able to focus particles to a particular point. Note that for the same particle speeds, in this figure $\beta=0.1$,  the magnitude of the deflections differ by a maximum value at their peaks. For $\beta=1$,  the maximum deflection is $\sim 4.6$~arcsec.  At about $r=0.075 R_\odot$, at the point of crossing-over of  power-law indexes $n=0$ to -2 for the density profile of the Sun (see text in \S 4), the difference in values of the deflection angles is small, suggestive of the fact that the Sun is a centrally condensed lens.}
\label{fig:1}
\end{center}
\end{figure}

The general formula for deflection in classical relativity for particles with arbitrary velocity $v$ is~\citep{accioly02}:
\begin{equation}
\delta_c=\frac{2 G M}{b c^2 \beta^2}\left(1+\beta^2+b g \left(1+\frac{\beta^2}{4}\right)\frac{3\pi}{2}+b^2 g^2\left(5 \beta^{-2}-\frac{1}{3} \beta^{-4}+\frac{5}{3}\beta^{2}\right)+\mathcal{O}(g^3)\right),
\end{equation}
where $g=GM/b^2$ and $\beta=v/c$. Keeping only the first order terms in $g$, the deflection is 
\begin{equation}
\delta_c=\frac{4 G M}{b c^2 }\left(\frac{1+\beta^2}{2\beta^2}\right).
\label{class-def}
\end{equation}
Note that in the limit $\beta\rightarrow 1$, we recover equation~(\ref{def}) and when $\beta\ll 1$ the amount of deflection is reduced by a factor of 2, which is essentially the Newtonian prediction~\citep{soldner}. 
The deflection angle obtained using the semi-classical approach, that is, by including  terms corresponding to the interaction of a massive photon field with a minimally coupled and static (external) gravitational field in the action integral, is~\citep{accioly04}
\begin{equation}
\delta_{sc} = \frac{4 G M}{b c^2 }\left(\frac{3-\beta^2}{2 \beta^2}\right).
\label{sc-def}
\end{equation}
In this case also, in the limit $\beta\rightarrow 1$, we recover equation~(\ref{def}). But, when $\beta\ll 1$, $\delta_{sc}=1.5 \delta=3 \delta_c$. For very slow-moving particles, the semi-classical result predicts a deflection 50\% more than the one obtained using the standard formula, equation~(\ref{def}). 
The values of deflection angles enter in the lens equation, the solution of which are the image locations for a given point-like source location.
The variation of the deflection angle as a function of particle speeds and impact parameters are shown in Fig.\ref{fig:1}.

In the following section, we will use two different deflection formulas to compute the minimum focal length for particles with a given velocity. The main reason for doing this is that because no one knows for sure which deflection formula provides the most accurate trajectory for slow-moving particles in the intermediate range of $\beta$ values, say from $10^{-3}$ to 0.1 (to be discussed in \S4), for it to be focused at minimum focal length(s) of 1~AU.

\section{\label{sec:3} Particle velocity and Focal Length}
Gravitational lensing by the transparent Sun generates a series of focal lengths starting with the minimum focal length of $\sim 23.5$~AU for neutrinos or any other fast-moving (non-massive) particles with velocities nearly equal to the speed of light. 
An approximate value  of the minimum focal length may be obtained by considering the mass enclosed within the transition impact parameter, at which the power-law index($n$) of  the density profile of the Sun changes roughly from $n=0$ to $n=-2$. The approximate formula for the minimum focal length for the Sun as given in ~\citet{patla08} is 
\begin{equation}
F\sim2\left(\frac{1+n}{3+n}\right)\left(\frac{c^2}{4g_\odot}\right)\left(\frac{r_n^2}{m_n}\right),
\label{appfoc}
\end{equation}
where $g_\odot=GM_\odot/R_\odot^2$, $r_n=0.075$ is the normalized radius corresponding to the transition of the power-law index and $m_n=0.1$ is the normalized mass enclosed within $r_n$, yielding a value of $\sim21$~AU. A higher limit for the  minimum focal length may be obtained directly from equation~(\ref{def}) instead, by considering the mass $m_n$ within an impact radius $r_n=0.075$, yielding a value of $\sim 31$~AU.

The extension of this formalism by using equations~(\ref{class-def}) and (\ref{sc-def}), corresponding to the classical and semi-classical deflection angles, and demanding that the approximate minimum focal length be $\sim 1$~AU imposes upper and lower bounds  for the speeds of particles that are subjected to the lensing action of the Sun.
$\beta$ values of $\sim 0.15$ and $\sim 0.26$, and 
$\sim 0.13$ and $\sim 0.22$ corresponding to minimum focal lengths of 20.5~AU and 31~AU respectively are obtained for both classical and semi-classical cases. 

Therefore, from rough estimates, we have to simply conclude that the velocities of slow-moving particles has to be within the range of 0.1 to $0.3 c$, in order to be focused on the Earth by the lensing action of the transparent Sun. Particles with speeds lesser than $0.1 c$ will have a series of focal lengths starting with a minimum value of less than 1~AU and particles with speeds greater than $0.3 c$ will be focused beyond the Earth's orbit.

This range of the velocity distribution will serve as good starting points for computing the actual focal lengths and magnifications in the following sections. For completeness, below we provide the relationship between the approximate focal length and the velocities of slow-moving particles using both classical and semi-classical formula (for deflections) respectively:
\begin{equation}
D_L=F_{\rm min}\left(\frac{2\beta^2}{1+\beta^2}\right),
\label{class-f}
\end{equation}
\begin{equation}
D_L=F_{\rm min}\left(\frac{2\beta^2}{3-\beta^2}\right),
\label{semi-f}
\end{equation} 
where $D_L$ is the distance between the lens and the observer and $F_{\rm min}$ is the minimum focal length for particles with speed $\beta=1$. Letting $D_L=1$~AU and solving equations~(\ref{class-f}) and (\ref{semi-f}) for $\beta$ yields the limiting velocities for the classical and semi-classical approximations for deflection angles.

The exact value of the limiting velocities of the particles that converge at 1~AU can be computed numerically, by using the deflections given by either of the formulas, equations~(\ref{class-def}) or (\ref{sc-def}), in place of equation~(\ref{def}) in the ray-tracing algorithm. The projected mass density is obtained by integrating the interpolated density profile of the Sun using the Standard Solar Model~(SSM) data published in~\citet{bahcall05}. The criterion employed for establishing the minimum focal length is the existence of an Einstein ring---the circle formed by the extended arcs of two images when the source, lens and observer are perfectly aligned in a straight line---the same criterion used in~\citet{patla08}. In depth discussions on the topics of Einstein rings, caustics and critical curves have been covered extensively in the literature on gravitational lensing~\citep{chwolson24,schneider92,nemiroff93,kochanek01}. We provide a brief and contextual overview of the same in \S5.

The exact value of the limiting particle speeds, using classical deflection formula given by equation~(\ref{class-def}) and semi-classical deflection given by equation~(\ref{sc-def}), corresponding to a minimum focal length of 1~AU, are  $(0.145\pm 0.001)c$ and $(0.247\pm 0.001)c$ respectively. 
Therefore, we cannot have a higher flux of particles with speeds higher than the values given above. 
The angular Einstein radii corresponding to the limiting speeds at 1~AU are given in Table~\ref{tab:1}. At 1~AU, the radius of the Sun measures up to 960~arcsec from Earth. 

\begin{deluxetable}{lcccccl}
\tabletypesize{\scriptsize}
\tablecaption{Particle speeds and focusing parameters for two deflection formulas }
\tablewidth{0pt}
\tablehead{
\colhead{$\delta\cdots\cdots\cdots\cdots\cdots\cdots\cdots\cdots\cdots\cdots\cdots\cdots$} & \colhead{$\beta~(v/c)$} & \colhead{$F_{\rm min}$~(AU) } & 
\colhead{$\theta_e$~(arcsec)} & \colhead{$\beta_c/\theta_e$} & \colhead{$\theta_c/\theta_e$} & \colhead{$M_{\rm max}$}
}
\startdata
$\delta_{c}$ & $ 0.145\pm 0.001$ & 1.00 & 41.2 & 0.0283 & $37.2\times10^{-2}$ & $\sim 10^6$\\
$\delta_{sc}$ & $ 0.247\pm 0.001$ & 1.00 & 38.3  & 0.0275 & $30.0\times10^{-2}$& $\sim 10^6$\\
$\delta_{c}$ & $0.100$ & $0.467\pm 0.001$ & 63.7   & 0.0315 & $32.2\times10^{-3}$ & $\sim 10^5$\\
$\delta_{sc}$ & $0.100$ & $0.160\pm 0.001$ & 221  & 0.0277 & $18.9\times10^{-2}$ & $\sim 10^5$\\
\enddata
\label{tab:1}
\end{deluxetable}

Also, as described in the caption of Fig.~\ref{fig:1}, the formalism of weak lensing fails for particle speeds that are close to and lesser than $\sim 0.01 c$, because the deflection values are too large to cause any focusing. But the trajectories of individual particles can be computed fairly accurately as long as the potential, $\phi \ll c^2$, submitting to the requirements of the weak field limit.

\section{\label{sec:4} Amplifications and flux}

The solution to the lens equation 
\begin{equation}
\vec{\beta_s}=\vec{\theta_i}-\vec{\alpha},
\label{lens}
\end{equation}
where $\vec{\beta_s}$ and $\vec{\theta_i}$ are the angles subtended by the unlensed source and its image(s), and $\vec{\alpha}$  is the deflection measure given by equations~(\ref{def}), (\ref{class-def}) or (\ref{sc-def}). The deflection values as a function of impact radius is obtained by projecting the mass enclosed within the impact radius onto the lens plane following the formalism developed in \citet{patla08,patla-phd}. The solution to equation~(\ref{lens}) are the image locations---roots of the polynomial equation---for any given source location.
For an introduction and history of gravitational lensing in the weak field limit, we refer the reader to the monograph by~\citet{schneider92} and/or  excellent reviews by \citet{blandford, narayan96,wambsganss}. 

We assume the Sun to be  a centrally condensed spherically symmetric lens. As a source crosses radially inward toward the  axis of the lens in the source plane, the number of images of the source changes from one to three in the lens plane~\citep{burke-bk}. The locus of all such points in the lens plane at which two images suddenly appear is a circle called critical curve. The area of the source plane comprising  source locations which corresponds to three images in the lens plane is encircled by a curve called the radial caustic. The term radial is used to suggest the images move radially apart from the caustic.
At the center of this radial caustic is a tangential point caustic; the corresponding tangential critical curve in the lens plane is the Einstein ring. At the tangential caustics the images stretch tangential to the caustic
The Einstein ring separates the new and old image sets in the lens plane. 
Magnifications are high along the critical curves  and caustics in the lens- and source-planes respectively as a result of the newly formed images.

In ordinary lensing germane to photons, the value of the Einstein ring (as a function of focal length) starts from an infinitesimally small value at the minimum focal length followed by a steep rise before attaining a maximum value to fall off gradually thereafter. For slow-moving particles, just like photons the value of Einstein ring as a function of focal length will mimic the pattern of the deflection angles that are plotted in panel~(b) of  Fig.~\ref{fig:1}. But the actual values, however,  will be different owing to the functional form of $f(\beta)$-terms in equations~(\ref{def}) and (\ref{class-def}).

\begin{figure}[htb]
\begin{center}
\includegraphics[width=\textwidth]{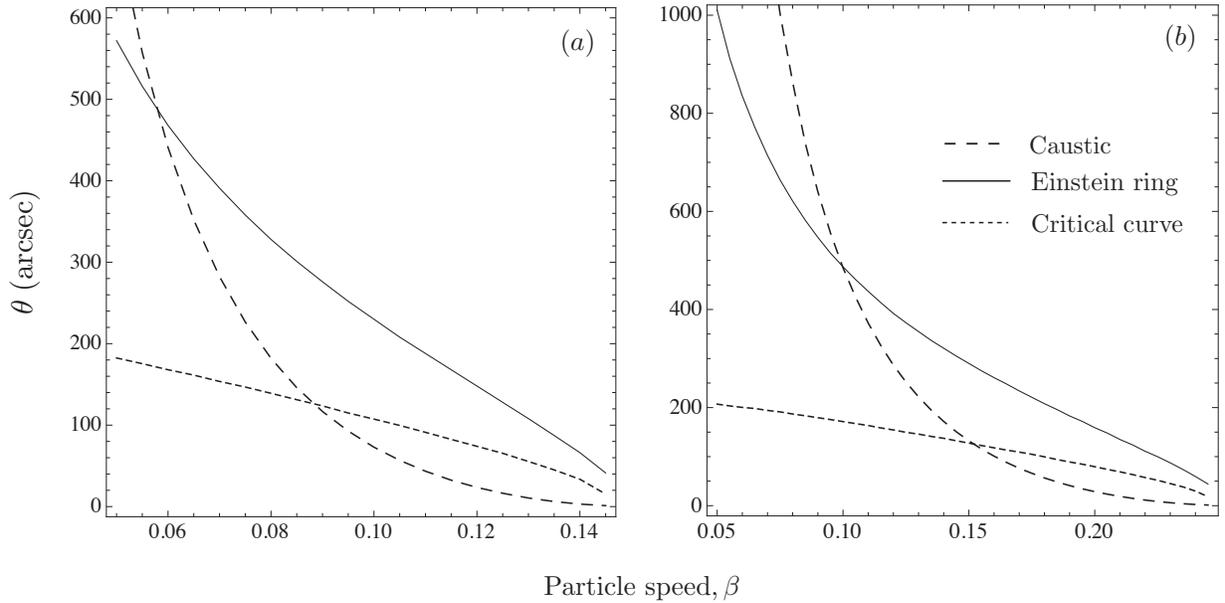}
\caption{Einstein radius, the radii of caustic and critical curves as a function particle speed for observer at 1~AU for (a) Classical and (b) Semi-classical deflection formulas. The images are always separated by the Einstein ring, but at lower speeds the caustics become larger than the critical curves--suggesting large deflections. For reference, at 1~AU, the radius of the Sun is 960~arcsec. The weak-field deflection limit breaks down when the value of Einstein ring becomes infinite.}
\label{fig:2}
\end{center}
\end{figure}

In other words, for slow-moving particles the radii of Einstein ring, caustics and critical curves will be larger if the observer is at or near to the minimum focal length when compared to lensing of neutrinos $\beta\approx 1$, for example. The dependence of particle speeds on the geometrical parameters of the problem is plotted in Fig.~\ref{fig:2}. 

Magnification is  defined as the ratio of the sizes of the image to that of the source and for a point-like source it is given by the formula $\mu=\theta d\theta/\beta_s d\beta_s$.  Very large magnifications of point-like sources occur when the source crosses a caustic. 
Magnification of an extended sources is computed by averaging individual points comprising the extended source.  Since all sources are small compared to the lens, our  formula for magnification is accurate enough for small impact parameter crossing of point sources across the lens. Moreover, perfect alignment of a point-like  source, the lens, and an observer along the lens axis is rather an exception than the norm.

On the question of how much magnification is possible, the answer is as high as$\sim 10^{6}$, not very different from the results of ~\citet{patla08}. Similar treatments concerning the magnification of point sources due to gravitational lensing, may be found in the excellent review article by~\citet{pacz96}. We use point-like sources for simulations involving the source crossing the lens at different impact parameters. Like in the previous sections, we use two different functions for deflection angles to solve for the lens equation. The roots of which are used for mapping regions of varying magnifications and for calculating the maximum values of magnification.

\begin{figure}[htb]
\begin{center}
\includegraphics[width=\textwidth]{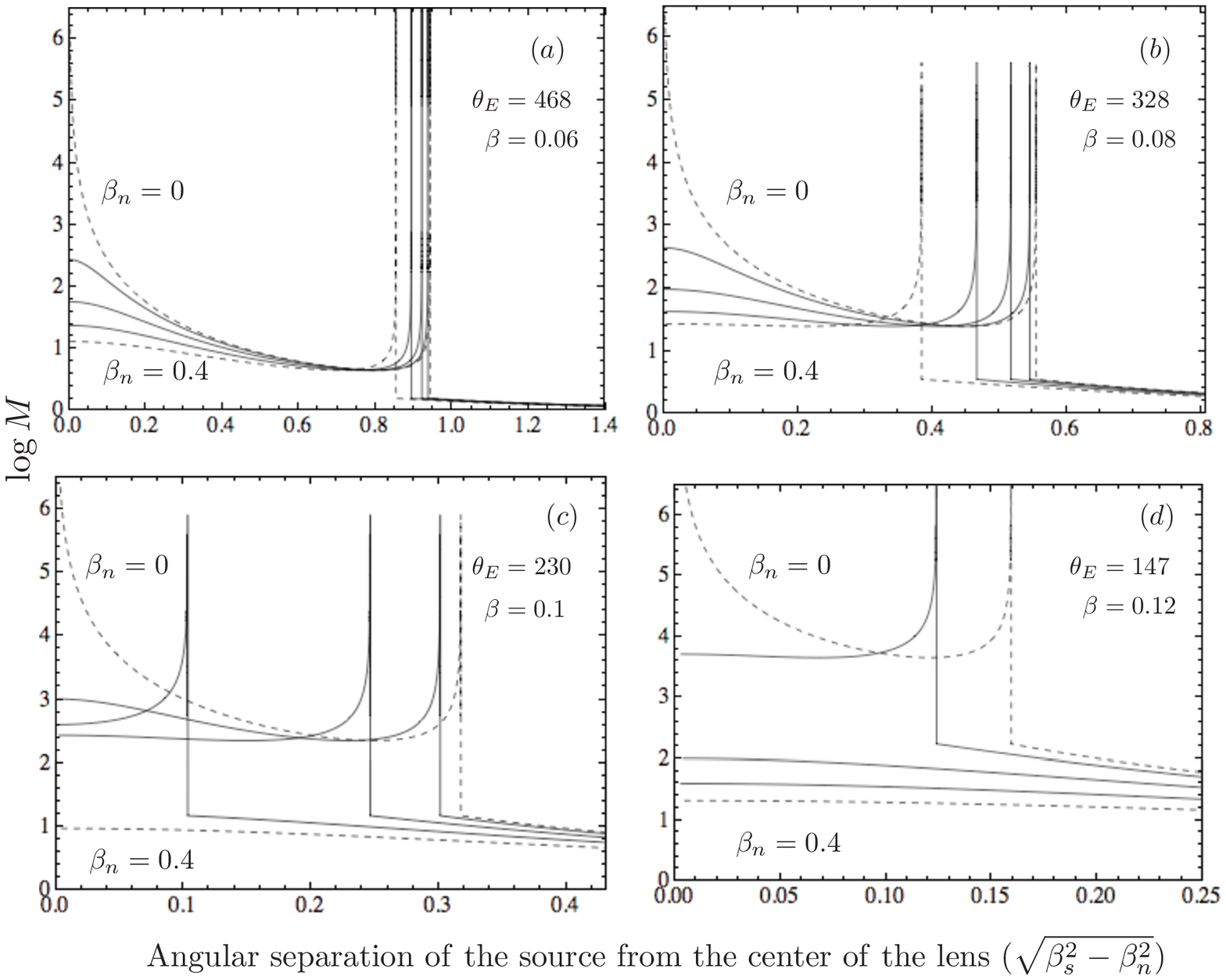}
\caption{Magnification for source crossing with impact parameters $\beta_n=0$ through $\beta_n=0.4$. $\beta_s$ and $\beta_n$ are normalized with the value of Einstein radius $\theta_E$ in units of arcsec.
Panels~$(a)$ through $(d)$ represent increasing particle speeds, $\beta=0.06$ through 0.12. The dashed lines represent maximum and minimum magnifications as the source is moving away from the radial and  toward the point caustic. The maximum magnification of $\sim 10^6$ is at the crossing of the radial($\beta_s=\beta_{\rm caus}$) and tangential($\beta_s=0,\beta_n=0$) caustics. For a finite source, realistic values of magnifications are $\sim 10^5$ with peak amplification periods of $\sim 0.1$~arcsec. Note that the radius of the Sun subtends as angle of 960~arcsec on Earth. The formula for classical deflection, equation~(\ref{class-def}) was used for generating deflection values for solving the lens equation. Note on conversion: 100 arcsec $\approx$ 30~min}
\label{fig:3}
\end{center}
\end{figure}

In Fig.~\ref{fig:3}, we present the magnifications for particles with speeds starting with $\beta=0.06$. For particles with lower speeds the magnifications are not as high and particles moving with a speed of $\sim 0.01c$ or less undergo way too much deflections and therefore cannot be focused. As the source moves across the lens, for a fixed impact parameter, the number of images multiply as the source gets closer to the center of the lens. The magnifications are high at or near radial and point caustics. 

For particles with intermediate range of speeds, $\sim 0.05c $ to $\sim 0.15c $ there will always be a significant amount of magnification over the radius of the caustic---distance separating the circumference of the radial caustic and tangential point caustic --- as evident from Fig.~\ref{fig:3}. The magnifications using the deflection formula of equation~(\ref{sc-def}) also yields similar results. The differences being, the particle speeds corresponding to maximum magnification will be shifted toward the right and the maximum magnifications will be close to an order of magnitude higher than the classical case as given in Fig.~\ref{fig:3}.




\section{\label{sec:5} Constraints on the mass of the slow-moving particle}
Just as we imposed velocity restrictions for recording high particle flux densities at the detector (starting with the Earth as the minimum focal point), in this section we use diffraction criteria to 
put limits on the particle mass in order to record maximum flux at the detector.

\begin{figure}[htb]
\begin{center}
\includegraphics[width=\textwidth]{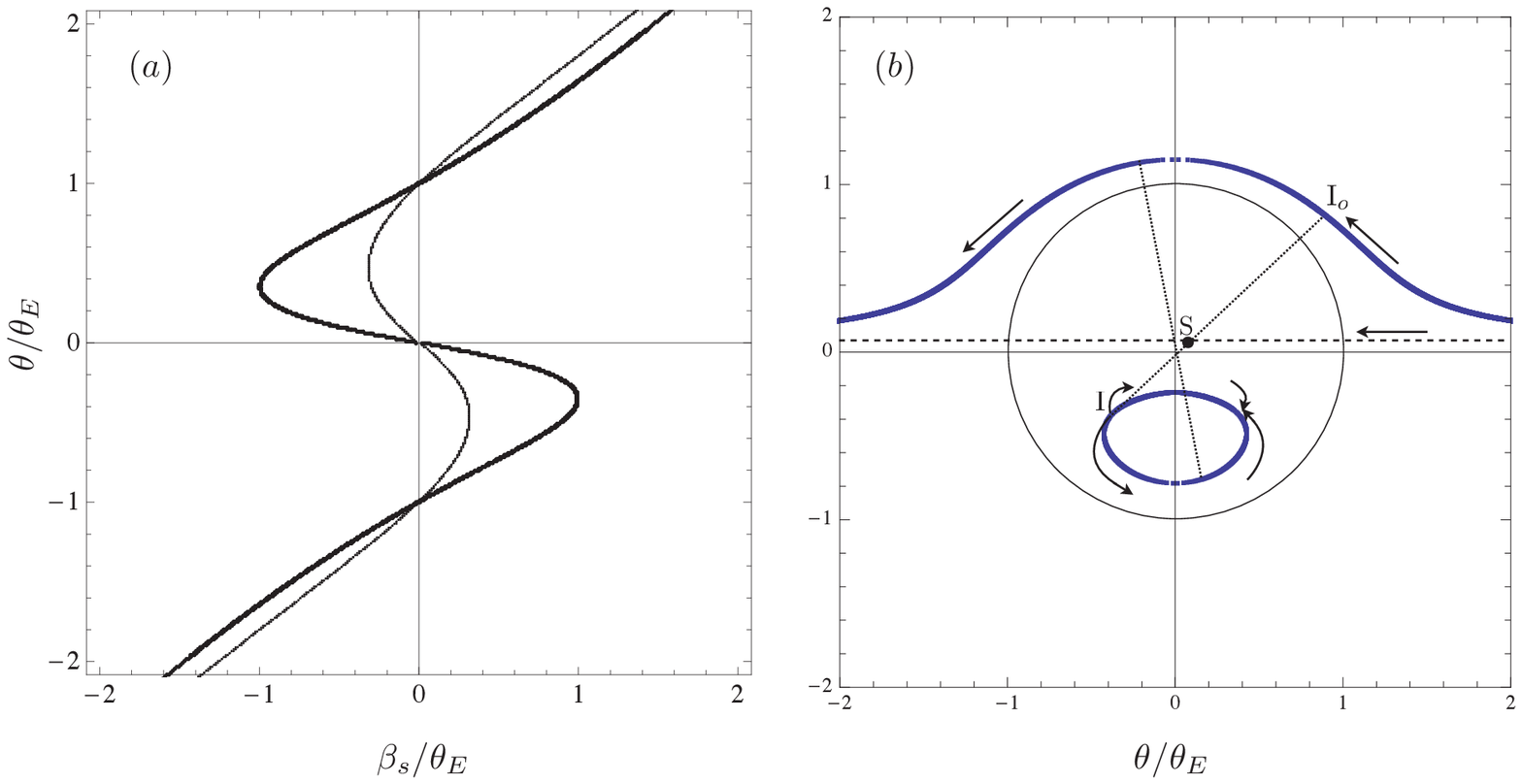}
\caption{$(a)$--Image locations as a function of source position, $\beta_s$. The thick curve represents the roots of the lens equation with the semi-classical deflection angle and the thin curve corresponds to the classical formula given by equation~(\ref{class-def}).The lens and observer is separated by a distance of 1~AU and the assumed particle speed is $\beta=0.1$. The semi-classical deflection formula yields a slightly higher magnification at the caustics owing to a larger deflection for the same particle speeds. $(b)$. Same as $(a)$, but the image formation illustrated in the lens plane. As a point source $S$ moves in from the right (along the dotted line), two images appear at $I$, then separates away from each other before merging again as the source exits  from the left. When a finite size source crosses the center of the lens, an Einstein ring (solid circle) is formed around the lens.}
\label{fig:4}
\end{center}
\end{figure}

The amplified image of a source (assumed of size smaller than the lens), envelops the Einstein ring when the source is perfectly aligned with the lens. For a point source, two point images are cast inside and outside of the Einstein ring as shown in Fig.\ref{fig:4}. 
At the observer location, such a source would be detected only if the individual particles from the source that appear to be coming from different points(images) in the lens plane would produce an interference pattern with a finite width ($w$) at the detector. The magnification is then roughly the ratio of the radius of the Einstein ring ($R_E$)  to the fringe width, $\mu \sim R_E/w$ \citep{ohanian,nakamura98,patla08}.

If the detector is at a distance $D_L$ from the lens, the Einstein radius is 
\begin{equation}
R_E=\left(\frac{4GM(b)D_L}{c^2 f(\beta)}\right)^{1/2}
\label{einsrad}
\end{equation}
where $b$ is the impact radius of the source and $f(\beta)=2\beta^2 (1+\beta^2)^{-1}$ or $2\beta^2 (3-\beta^2)^{-1}$ depending on the deflection formula used to compute the image locations. Noting that $w\sim \lambda D_L/R_E$, the maximum magnification is
\begin{equation}
\mu\sim \frac{R_E^2}{\lambda D_L}=\frac{2GM(b)}{c^2} \frac{2}{\lambda f(\beta)} \sim \frac{R_S}{\lambda f(\beta)},
\label{sch-eins}
\end{equation}
where $R_S\sim 3$~km is the Schwarzschild radius of the Sun.
Therefore, when $\beta=1$, the condition for large magnification factors is $R_S \gg\lambda$ and for $\beta<1$ this condition may be relaxed by a factor $\sim\beta^{-2}$. Assuming the wavelength $\lambda$ for slow moving particles of mass $m_0$ to be its de~Broglie wavelength $\lambda \approx\lambda_{dB}=h/(\gamma m_0 \beta c)$, where $\gamma=(1-\beta^2)^{-1/2}$,
\begin{equation}
\lambda_{dB}=\lambda_c \left(\frac{1-\beta^2}{\beta^2}\right)^{1/2},
\label{debro}
\end{equation}
where $\lambda_c=h/m_0 c$ is the Compton wavelength of the particle. For appreciable magnification at 1~AU, the particles must have a mass of at least
\begin{equation}
m_0\ge\frac{h}{c}\left(\frac{1}{\theta_E^2 D_L}\right)\left(\frac{1-\beta^2}{\beta^2}\right)^{1/2}.
\end{equation}
Using the values for $D_L=1$~AU from Table~\ref{tab:1}, the minimum value for $m_0=1.68\times10^{-45}$~kg. Alternatively, using the same mass fraction ($m_n=0.1$) from  the approximate formula given by equation~(\ref{appfoc}), we approximate the limiting mass of the particle
\begin{equation}
m_0(\beta)=\frac{hc}{4G  M_\odot m_n} \left(\frac{1-\beta^2}{\beta^2}\right)^{1/2} f(\beta),
\end{equation} 
which gives a minimum mass of $m_0=3.0\times 10^{-46}$~kg. Taking the higher (and also most accurate) value of the limiting mass yields a particle mass of at least $10^{-9}$~eV, which covers almost all dark matter particles considered so far.

As a instructive exercise, from Table~\ref{tab:1}, consider the semi-classical case deflection (just because its value is higher) at 1~AU. If indeed the prediction of a narrow energy window for axions as explained in \S 2 holds true, the above diffraction constraints allow up to a million $10^{-4}$~eV (just as an example) particles confined to an area of $\sim 100~\rm{m}^2$. Similar numbers may be computed for various particle speeds using equations~(\ref{einsrad}) through (\ref{debro}) using values for corresponding variables from Fig.\ref{fig:3}.

\section{\label{sec:6} Lensing by planets: Jupiter as an example}
Large planets like Jupiter are also capable of lensing slow-moving particles. In this section we consider 
scenarios when planets could act as the primary lens or serve to further amplify the lens action of the Sun as they transit and perfectly align along the Sun-Earth axis.

\subsection{Planet as lens}
Jupiter is separated by a distance of $\sim 4.2$~AU and $\sim 6.2$~AU from its closest and farthest approach from Earth~\citep{jupnasa}. The internal density variation of Jupiter is not well understood; it's also not clear whether Jupiter even has a core. However, studies have attributed it to having a core that might be as large as $0-18 M_\oplus$~\citep{guillot, nettelmann}. Unlike stars, one would not expect planets to be  centrally condensed. So to keep it general, we will develop a formalism involving a density profile that is more representative of planets. For now, we will introduce a free parameter (core radius) that could be fine-tuned to fit the actual Jovian density profile when it becomes available.   

\begin{figure}[htb]
\begin{center}
\includegraphics[width=0.6\textwidth]{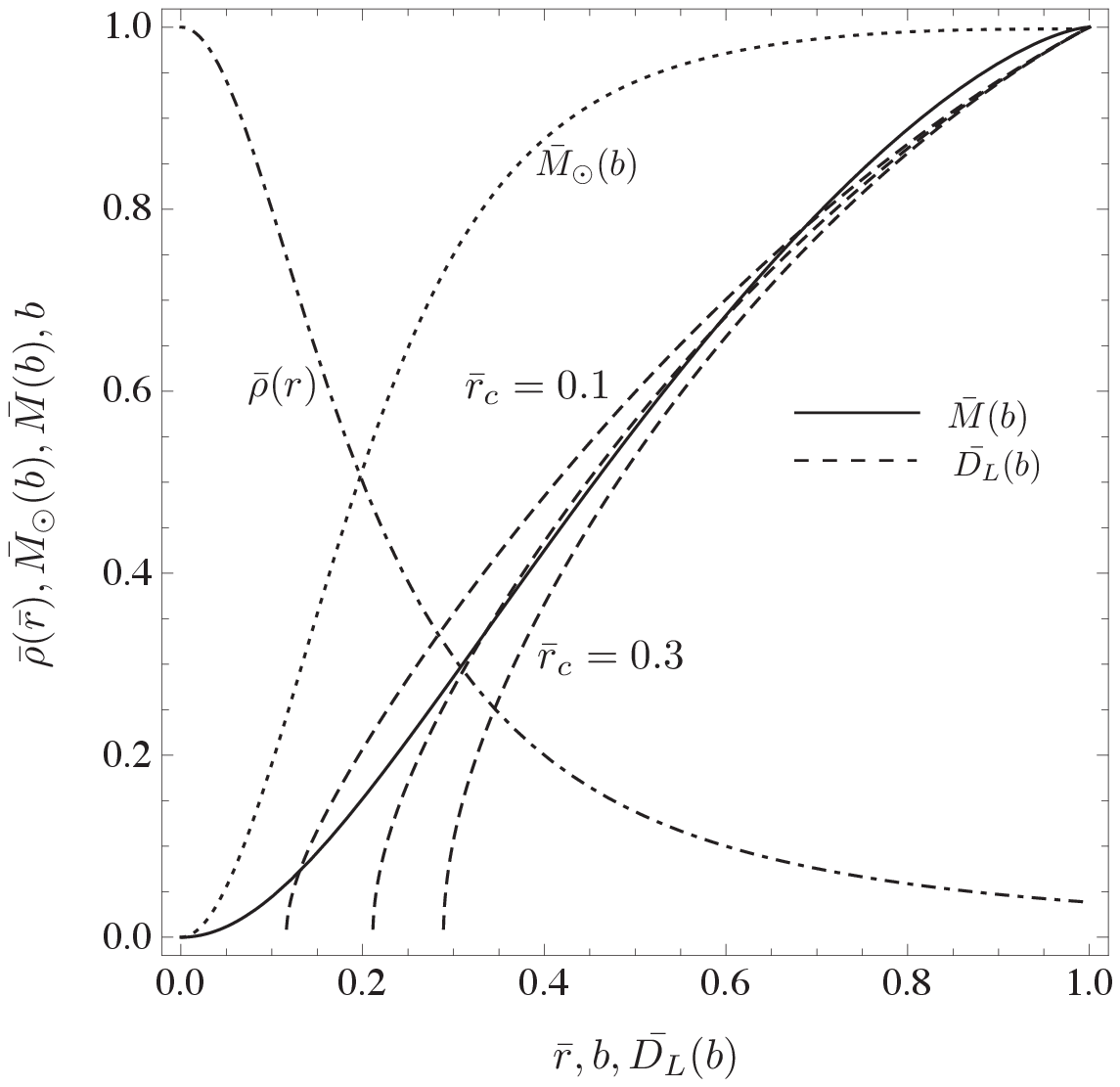}
\caption{Normalized density, projected mass and focal length (see equations~(\ref{plden}),(\ref{plmassenc}) and (\ref{plfoc})) for Jupiter as functions of dimensionless impact parameter $b$. Axes for the plots of focal lengths corresponding to three different values of normalized core radii ($\bar{r}_c=0.1,0.2$ and $0.3$) are reversed for the convenience of directly inferring the values of minimum focal lengths for validating equation~(\ref{plminfoc}). For comparison, the projected mass of the Sun from the SSM data---indicative of a centrally condensed lens---is also shown alongside with the projected mass from an assumed density profile for Jupiter.}
\label{fig:5}
\end{center}
\end{figure}

We'll assume a Lorentzian density profile with a core radius for approximating a more general planetary lens:
\begin{equation}
\rho(r)={\rho_0\over 1+\left(\frac{r}{R_c}\right)^2},
\label{plden}
\end{equation}
where $R_c$ is the assumed radius of the core and 
\begin{equation}
\rho_0=\frac{M_\star}{4\pi R_{c}^3}\left[\frac{R_\star}{R_c}-\tan^{-1}\left(\frac{R_\star}{R_c}\right)\right]^{-1}.
\label{pldencon}
\end{equation}
$M_\star$ and $R_\star$ are the mass and the radius of the planet. In order to obtain a minimum focal length for small impact radii, we first obtain the projected mass as a function of impact radius
\begin{equation}
M(r)=2\int_0^{2\pi}\int_0^{r}\int_0^{\sqrt{R_\star^2-R^2}}\rho(r)dRdz,
\label{plmassenc}
\end{equation}
where $r^2=R^2+z^2$. Using appropriate substitutions and performing integration by parts of the inner two integrals, we obtain the projected mass in terms of the dimensionless impact parameter
\begin{equation}
\frac{M(b)}{M_\star}=\frac{f(b)-f(0)}{r_c}\left[\frac{1}{r_c}-\tan^{-1}\left(\frac{1}{r_c}\right)\right]^{-1}\quad {\rm for }\quad  b\le 1
\end{equation}
where,
\begin{equation}
f(\bar{r})=\sqrt{r_c^2+\bar{r}^2}\left[\tan^{-1}\left(\frac{1+r_c^2}{r_c^2+\bar{r}^2}-1\right)^{1/2}-\left(\frac{1+r_c^2}{r_c^2+\bar{r}^2}-1\right)^{1/2}\right],
\end{equation} 
$\bar{r}\equiv r/R_\star$, $r_c\equiv R_c/R_\star$ and  $b\equiv r/R_\star$. For $b>1$, $M(b)=M_\star$. The focal length is 
\begin{equation}
D_L(b)=\frac{b^2 c^2 f(\beta) R_\star^2}{4 G M(b)}.
\label{plfoc}
\end{equation}
In order to obtain the minimum focal length, we take the limiting case for small impact parameters,
\begin{equation}
\lim_{b\rightarrow0} D_L(b)\equiv F_{\rm min}=\frac{c^2 f(\beta)}{4g_\star}\frac{2 r_c^2}{\tan^{-1}\left(\frac{1}{r_c}\right)}
\left[\frac{1}{r_c}-\tan^{-1}\left(\frac{1}{r_c}\right)\right],
\label{plminfoc}
\end{equation}

where $g_\star=G M_\star/R_\star^2$.
Now specifically for Jupiter, $g_\star\sim~25~$m/s$^2$. Letting $r_c=0.2$ roughly satisfies the 
criterion that $0-18 M_\oplus$ is contained within a radius $R_\oplus$ comprising Jupiter's core. Substituting the value of $r_c$ in equation~(\ref{plminfoc}) yields a minimum focal length of $\sim$~1200~AU. Depending on our knowledge of Jupiter's core, which we hope will become available in the future, the value of $r_c$ may be tweaked appropriately to match with the data. The focal length corresponding to the Jovian limb as the impact radius is $\sim6000$~AU, compared to $\sim 550$~AU corresponding to the  Solar limb. The values for focal lengths corresponding to a range of values for the core radius are given in Fig.\ref{fig:5}.

Substituting the values for $D_L$ as 4.2 and 6.2~AU in equations~(\ref{class-f}) and (\ref{semi-f}) yields a particle velocity of $0.04c$ and $0.05c$ for classical, and $0.07c$ and $0.08c$ for semi-classical cases respectively. As in Fig.\ref{fig:1}, the maximum deflections may be computed for Jupiter for maximum deflections corresponding to $r_c=0.2$. The pattern of the curves resemble with the ones obtained for the Sun, except for the fact that the lower limit for particle speed is now an order of magnitude less compared to Sun. Therefore, at least with the assumption of $r_c=0.2$, Jupiter is capable of focusing particles with speeds as low as $0.001c$. This fact is noteworthy, because by most estimates the detectable dark matter spectrum on Earth peaks at $0.001c$~\citep{sikivie}.

The magnifications are expected to be as high as the values obtained for the Sun, at least near the caustics. We will postpone the study of caustics, event separations, etc. to some time in the future, as the density profile of Jupiter, at this time, is very much uncertain. Also neglected in our analysis  are the many moons of Jupiter.

\subsection{Transiting Jupiter behind the Sun}

The effect of transiting massive planets like Jupiter, when aligned perfectly with the Sun, would alter the minimum focal length. Assuming a mass of $\sim 10^{-3}~M_\odot$ for Jupiter, for particle speed(s) corresponding to the minimum focal length of 1~AU given in Table~\ref{tab:1}, the reduction in focal length, using the approximate formula given by equation~(\ref{appfoc}), is $\sim 0.01$~AU. The effect of which leads to an increase in the value of the Einstein radius observed at 1~AU and therefore, a slight enhancement in magnification. The Einstein radii, for the Sun, as a function of focal length rises to a maximum within roughly twice the value of the minimum focal length and falls off monotonically~\citep{patla08}. Since the reduction in minimum focal length is less than a percent as a result of Jupiter's transition, the accompanying increase in magnification will be less than an order of magnitude.

However, slightly higher velocity particles that wouldn't previously be focused or focused with reduced flux at 1~AU, will now be detectable at this location. Also, the separation between the magnification peaks for the same source with a fixed transiting speed and impact parameter will also be increased by a few percent (compared to the case of a planet not aligned along with the lens axis).

\section{\label{sec:7} Discussion}

Although not much is known about what objects are capable of generating an appreciable supply of slow-moving particles, our Sun is suggestive for likely sources which would include  massive stars that not only dissipate energy continuously, but also transiently while undergoing dynamic transformations: implosions or explosions. 
The sources could also include massive bodies that are capable of gravitationally attracting a flux of surrounding dark matter particles and reemitting them instead of producing them. For example, based on galactic halo formation models, \citet{sikivie} conclude that the dark matter particle abundance near the Earth peaks at velocities of the order of $\sim 0.001 c$, also see~\citet{griest}. It is plausible that galaxy centers can attract such fluxes and redirect them toward the Sun and thus make accessible to us the lower-end of the velocity
spectrum, which is roughly half the maximum.

If indeed strong isolated sources producing or reemitting slow-moving particles do exist, it is reasonable to assume an emergent flux that is centered around a mean velocity. To answer the question, what is the velocity resolution of the flux that can be inferred by a pair of detectors on Earth, we use the relationship between velocity difference  and focal length, obtained from the formula given by equation~(\ref{class-f}), 
\begin{equation}
\frac{\delta F}{F}=\frac{f'(\beta)}{f(\beta)}\delta\beta.
\label{resolution}
\end{equation}
This gives a maximum value for the velocity resolution $\delta\beta_{\rm max}\sim 10^{-5}\beta$ for particle flux with a given mean speed
$\beta$, assuming that two detectors are separated by a distance roughly (to first order in $\beta$) equal to the diameter of the Earth. The minimum value is set by the astigmatism of the Sun, which is of the order of few meters and is roughly $\delta\beta_{\rm min} \sim 10^{-10}\beta$~\citep{gerver}.

The magnification of the flux of particles coming from the source, as a result of the source straddling the lens plane, will peak only during the times when the source is near the caustics. Considering each peak as a lensing event, for a source at sufficiently small impact radii, the temporal separation between any two such lensing events (see  Fig.~\ref{fig:3}) is about an hour or less. This time scale is based on the motion of the Earth around the Sun. Moreover, most distant objects in the sky that are of interest to us move at much smaller speeds.

Based on the temporal separation of the lensing events and based on measured mean velocity, we could potentially rule out one of the deflection formulas given by equations~(\ref{class-def}) and (\ref{sc-def}). 
This is because when equation~(\ref{class-def}) is recast in terms of particle energies, the deflection becomes dependent on the total energy of the particle. In that case one might argue that such particles could violate the equivalence principle (EP)~\cite{accioly04}.

One related question then is could we possibly test for the violation of EP with  at least a select variety of particle species. If equation (\ref{class-def}) is ruled out then most likely the EP violation is a possibility. Alternatively, the lower limit of the velocity resolution of the particle flux might not be small enough to validate EP violation ($\delta\beta_{\rm min}/\delta t^2$).
Because the current understanding is that there is no violation reported for values of $\eta$ parameter (measure of relative acceleration for two different masses) up to  $\sim 10^{-13}$~\citep{patla12}. A thorough analysis is needed  for accurately designing an experiment to test EP using the lensing of slow-moving particles and so requires additional work.

 Invoking diffraction criteria, we constrain the mass of the slow-moving particles to be focused at the detector to be more than $\sim 10^{-9}$~eV. The Sun and Jupiter may focus slow-moving particles that comprises the bulk of the predicted dark matter spectrum on Earth. Jupiter has the potential to amplify flux here on Earth that the Sun is not capable of focusing: particles with speeds $0.01c-0.001c$. Also, the perfect alignment of  Jupiter behind the Sun will give a marginal flux enhancement amounting to an increase of less than an order of magnitude.  

We note that astigmatism of the Sun is of the order of a few meters and as a result the maximum magnification could be reduced by an order of magnitude~\citep{gerver}. The effect of Sun's relative motion in the galaxy and the resulting flux modulation has negligible effects for isolated point-like sources that we have considered~\citep{spergel_sun}.
Uniform and isotropic background sources produce no discernible magnifications, just like large size sources produce reduced magnifications. We, therefore, consider only point-like cosmological sources.

In conclusion, it  is possible for detecting the flux of slow-moving and  non interacting  particles lensed by the Sun or Jupiter---here on Earth. In order to be detected on Earth the particles have to have speeds between $\sim 0.01 c$ and $.14 c$ for the Sun. If semi-classical deflection angles are considered, these speeds will scale to $\sim 0.01 c$ and $.24 c$. Particles with speeds less than $\sim 0.01 c$ will undergo way too much deflection to be focused, although such individual particles could be detected. 
The same is true for Jupiter, albeit the values being scaled back by an order of magnitude.
The magnifications can be as high as $\sim 10^{6}$ at the caustics for point sources, with a more reasonable value of $\sim 10^{5}$ for real situations involving small sources crossing the caustics. Substantial magnification of $\sim 10^4$ is possible for times ranging up to 30~min, although peak amplifications of  $\sim 10^{6}$ prevail only  for about few seconds (0.1~arcsec) or even less.



\acknowledgments
BP thanks Lakshmi, Juno and Charlie for their patience and encouragement during the time spent writing 
this paper.

%



\bibliographystyle{apj} 
\bibliography{dmlens}    

\clearpage



\end{document}